# Ring like correlation in relativistic heavy ion collision: an experimental probe using continuous wavelet transform approach

Gopa Bhoumik, Argha Deb, Swarnapratim Bhattacharyya[*] and Dipak Ghosh

**Nuclear and Particle Physics Research Centre**
**Department of Physics**
**Jadavpur University**
**Kolkata - 700 032**
**India**

* **Department of Physics, New Alipore College,**
**L Block, New Alipore, Kolkata 700053, India**
Email: argha_deb@yahoo.com

## Abstract

Continuous wavelet transform approach has been applied to the pseudo-rapidity distribution of shower particles produced in $^{16}$O-AgBr interactions at 60 AGeV and $^{32}$S-AgBr interactions at 200 AgeV. Multiscale analysis of wavelet pseudo-rapidity spectra has been performed in order to find out the presence of ring-like correlation, which could be either due to production of Cherenkov gluons or due to propagation of Mach Shock wave through excited nuclear matter. This approach fulfils the basic requirement of both effects that they lead to an overabundance of considered particles at some typical pseudo-rapidities. Comparison of experimental results with that obtained from analyzing events generated by FRITIOF code are not reproduced.

## I. Introduction

The problem of presentation of high multiplicity events produced in high energy nucleus-nucleus interactions is non-trivial. The presentation of each particle in 3-dimensional phase space corresponds to a dot. These dots form different pattern in phase-space which may be linked to different dynamics of multiparticle production. The distribution of these dots can be found for an ensemble of events and for a single event also. The average distribution over all events possesses fractal properties **[1]** at low energies. To study the properties of nuclear matter that may have undergo some qualitative changes, in hot and compressed condition collective effects are of prime importance.

It is also interesting to study phase-space pattern for a single event. But for a single event statistical fluctuation may dominate in the observed effects. Dense particle clusters are observed at different positions and at different scales of the considered phase-space. Analysis of these clusters on an event-by-event basis is informative.

The factorial moment method **[2]** was proposed to remove the background statistical fluctuation in global analysis. Recently a new type of mathematical approach have been developed, the so-called wavelet analysis, which has been applied successfully in different fields of science and engineering **[3-6]**.The method is also fruitful in multiparticle data analysis as for a single event it can reveal the local characteristics of any pattern of particle distribution after eliminating the smooth polynomial trends at different scales. Thus one gets the opportunity to study strong dynamical fluctuation removing the statistical component. The wavelet analysis was first used in multiparticle data analysis by P. Carruthers **[7-9]** to diagonalise the covariance matrix of some simplified cascade model. Later several works using wavelet analysis on multiparticle production process have been done **[10-15]** at energy $E_{lab}$=10-10$^3$GeV/nucleon.

We have considered $^{16}$O-AgBr interactions at 60 AGeV and $^{32}$S-AgBr interactions at 200 AGeV. The pseudo rapidity distribution of the pions produced in all events of these interactions is analysed using continuous wavelet based approach and also we have selected one high multiplicity event from each considered interaction for analysis. Our main aim is to detect some conspicuous peaks or irregularities, which may reveal the preferred emission polar angle in multiparticle production process. Production of large number of particles at some distinguished pseudo rapidity values may indicate the presence of ring-like structure if relativistic particles are considered for analysis. So in this regard we hope wavelet analysis will be fruitful to characterise the η distribution at different scale to localize multiparticle correlation that may involve different numbers of particles. We have also compared our experimental results with that obtained from analysing events, generated using FRITIOF code.

The rest of the paper is arranged as follows: in section II a brief description of the experiment has been given. Section III describes in detail the wavelet analysis method. In section IV we have discussed our findings followed by a conclusive section (section V).

## II. Experimental Details

In the current analysis we have considered the interactions of the $^{16}$O beam, moving at energy 60A GeV, and $^{32}$S beam, moving at energy of 200 A GeV, with AgBr being the target present in a nuclear emulsion. Small stacks of ILFORD G5 emulsion plates were exposed to the above mentioned beam during EMU-08 experiments**[16-18]** at CERN. SPS accelerator used to accelerate the beams. The details of the experiment like: dimension of emulsion plate, beam flux have been given in our previous publication **[19]**.

For the scanning procedure of the emulsion plates containing interactions, we have used a Leitz Metalloplan microscope. The objective and ocular lens of this microscope has 10 times magnifying power with a semi-automatic scanning stage. In order to increase the scanning efficiency, each emulsion plate was scanned by two independent observers. The final measurements were carried out with the help of an oil-immersion objective of 100× magnification. The resolution of the measuring system is 1 $\mu$m along the beam direction (*X*-axis) and along *Y*-axis and 0.5 $\mu$m resolution along the *Z*-axis, which corresponds to the thickness of the emulsion plate. The projectile beam passes through the emulsion plate horizontally i.e. along X-Y plane. In nuclear emulsion terminology **[20]**, particles emitted after interaction are classified as the shower, gray and black particles. Shower particles comprises mostly (about more than 90%) of pions with a small admixture of K-mesons and hyperons. They have ionization $I \leq 1.4I_0$, where $I_0$ is the minimum ionization value, and velocity greater than 0.7 *c*. Grey particles are mainly fast target recoil protons with energies up to 400 MeV, having ionization $1.4I_0 \leq I < 10I_0$ and velocities laying between $0.3c$ and $0.7c$. Black particles consist of both singly and multiply charged fragments with ionization $I \geq 10I_0$ and velocity less than $0.3c$. Beside, these particles there could be also a few projectile fragments. They do not directly participate in an interaction and regarded as the spectator parts of the incident projectile nuclei that.

In selecting the events following criteria are used:

(a) The incident beam track should lay within 3° from the direction of the main beam in the pellicle. (b) Events, which show, interactions within 20 μm from the top and bottom surfaces of the pellicle were rejected. (c) All the suggested primary beam tracks were followed in the backward direction to exclude the events, arising from the secondsry tracks of other interactions.

In emulsion plate interaction of incoming beam may occur with three different types of targets, e.g. hydrogen (H), light nuclei (CNO) and heavy nuclei (AgBr) present in the emulsion medium. Let $N_h$ represent the total number of black and grey tracks, together called heavy tracks. Collision between hydrogen and the projectile beam give rise to events with $N_h \leq 1$. Events with $2 \leq N_h \leq 8$ occur due to collisions of projectile with light nuclei and events with $N_h > 8$ are due to collisions with heavy nuclei. In our study, in order to exclude the H and CNO events, we have considered only those events having number of heavy tracks greater than 8.

According to the above described selection procedure, we have chosen 250 events of $^{16}$O–AgBr interactions at 60 A GeV [21] and 140 events of $^{32}$S–AgBr interactions at 200 A GeV[22]. In the present analysis we considered the shower tracks only. The average multiplicities of the shower tracks are 63.20 ± 0.21 and 93.82 ± 0.18 in the case of $^{16}$O–AgBr and $^{32}$S–AgBr interactions, respectively. The emission angle ($\theta$) was measured for each track with respect to the beam direction by taking readings of the coordinates of the interaction point ($X_0$, $Y_0$, $Z_0$), coordinate ($X_1$, $Y_1$, $Z_1$) of the point at the end of the linear portion of the incident beam and ($X_i$, $Y_i$, $Z_i$) of a point on the incident beam. The pseudo-rapidity variable ($\eta = -\ln \tan \frac{\theta}{2}$), which may be treated as a convenient substitute of the rapidity variable of a particle when the rest mass of the particle can be neglected in comparison to its energy or momentum, has been determined for each pion track for further analysis.

### III. Wavelet Analysis Method

This is a practical mathematical tool which offers one to perform multiscale analysis of nonstationary or in homogeneous signal. The signal may comprise of ordered set of any numerical information recorded over any process, object or function. Wavelet construction of any signal consists of two parameters: i) dilation or scale parameter '$a$' and ii) translation or position parameter '$b$'. Tuning the scale parameter local analysis can be performed. Small '$a$' reveals the property of a single particle whereas larger '$a$' values resolve cluster of particles. On the other hand parameter '$b$' enables to scan the whole range of the signal. Unlike Fourier transform, which uses only two basic set of functions, wavelet analysis offers an infinite numbers of basic functions for discrete and continuous analysis. However one have to pick up the appropriate one according to basic features of the signal like: shape, regularity, symmetry, continuity etc.. In output the information's, extracted from the data, are more clear and interpretable.

In general the continuous wavelet transform of any function $f(x)$ takes the form **[23, 24]**

$$W_\Psi(a,b)f = \frac{1}{\sqrt{C_\Psi}} \int_{-\infty}^{\infty} f(x)\, \Psi_{a,b}(x)\, dx$$

(1)

where $C_\Psi$ is a normalizing constant and given by the following integration: $C_\Psi = 2\pi \int_{-\infty}^{\infty} |\widetilde{\Psi}(\omega)|^2 |\omega|^{-1}\, d\omega$ where $\widetilde{\Psi}(\omega)$ is Fourier transform of $\Psi(x)$ and $\Psi_{a,b}(x) = a^{-1/2}\, \Psi\left(\frac{x-b}{a}\right)$ is the shifted and/or dilated form of the mother wavelet function $\Psi(x)$. The wavelet-transform coefficients $W_\Psi(a,b)$ measures the contributions of $\Psi_{a,b}(x)$ to $f(x)$.

If the distribution of N experimentally measured values $x_i$ be expressed as

$$f(x) = \frac{dn}{dx} = \frac{1}{N}\sum_{i=1}^{N} \delta(x - x_i) \quad (2)$$

then the coefficients of continuous wavelet transform of (2) takes the form

$$W_\Psi(a,b) = \frac{1}{N}\sum_{i=1}^{N} a^{-1/2}\, \Psi\left(\frac{x_i - b}{a}\right)$$

In our present case $x_i$'s stand for the pseudo-rapidities $\eta_i$ of shower tracks of the studied sample (a single event or total ensemble).

As a choice of mother wavelets, the derivatives of the Gaussian function

$\Psi(x) = g_n(x) = (-1)^{n+1} \frac{d^n}{dx^n} e^{-x^2/2}$ are often used. For multiparticle emission data analysis, in this case we have selected the second derivative of the Gaussian function, well-known as Mexican Hat distribution (shown in fig. 1), $g_2 = (1 - x^2)e^{-x^2/2}$ as mother wavelet.

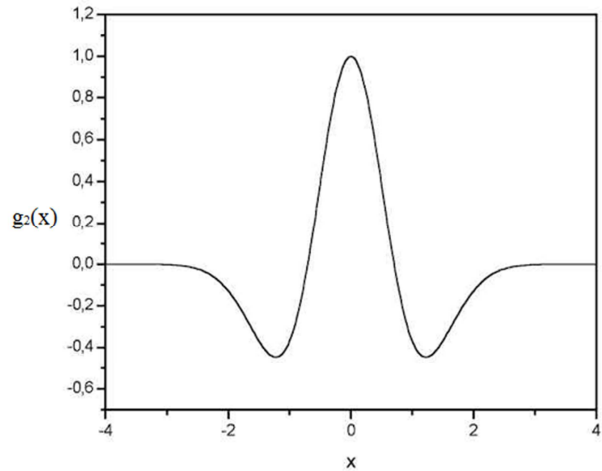

Figure. 1. Mexican hat wavelet.

The obvious reason for that is the Gaussian-like signal has been assumed. Another advantage of using $g_2$ wavelet is that it attains feasible resolution both in scale and pseudorapidity domain simultaneously. Using this algorithm we will find out the wavelet pseudorapidity spectrums, which are nothing but the sum of wavelets representing the individual particles.

## IV. Results and Discussions

Wavelet pseudo-rapidity spectra of all the events of the considered interactions will be presented first. In this approach one may find out the collective flow occurring in many events and not the unique behaviour of an individual event. In figure 2 we have presented the wavelet pseudo-rapidity spectra of the shower track distribution of $^{16}$O-AgBr interactions at 60 AGeV for different scales '$a$'. The events generated by FRITIOF code based on Lund Monte Carlo model **[25-27]**, can approximately reproduce the pseudo-rapidity distribution of the shower particles in the considered interactions **[28]**. So here, we have compared our results with that obtained from events generated by FRITIOF code. These results are shown in the same figure by the dotted line.

It is obvious from the figure that $g_2$ pseudo-rapidity spectra at small scale reveal an intricate fine structure, whereas for larger scale values only coarse features are observable.

The '$b$' values at the peaks of the spectra correspond to characteristic pseudo-rapidity values where particle clusters are formed. From the size of area under each peak one can get an idea about the number of particle in each group. It is observed that three regions of maximums are present there and they are interpreted as the central particle producing region ($\eta \sim 2.3$), at the left side of the central region ($\eta \sim 0.2, 1.2$) the target fragmentation region and that at the right side ($\eta \sim 3.7$) correspond to projectile fragmentation region.

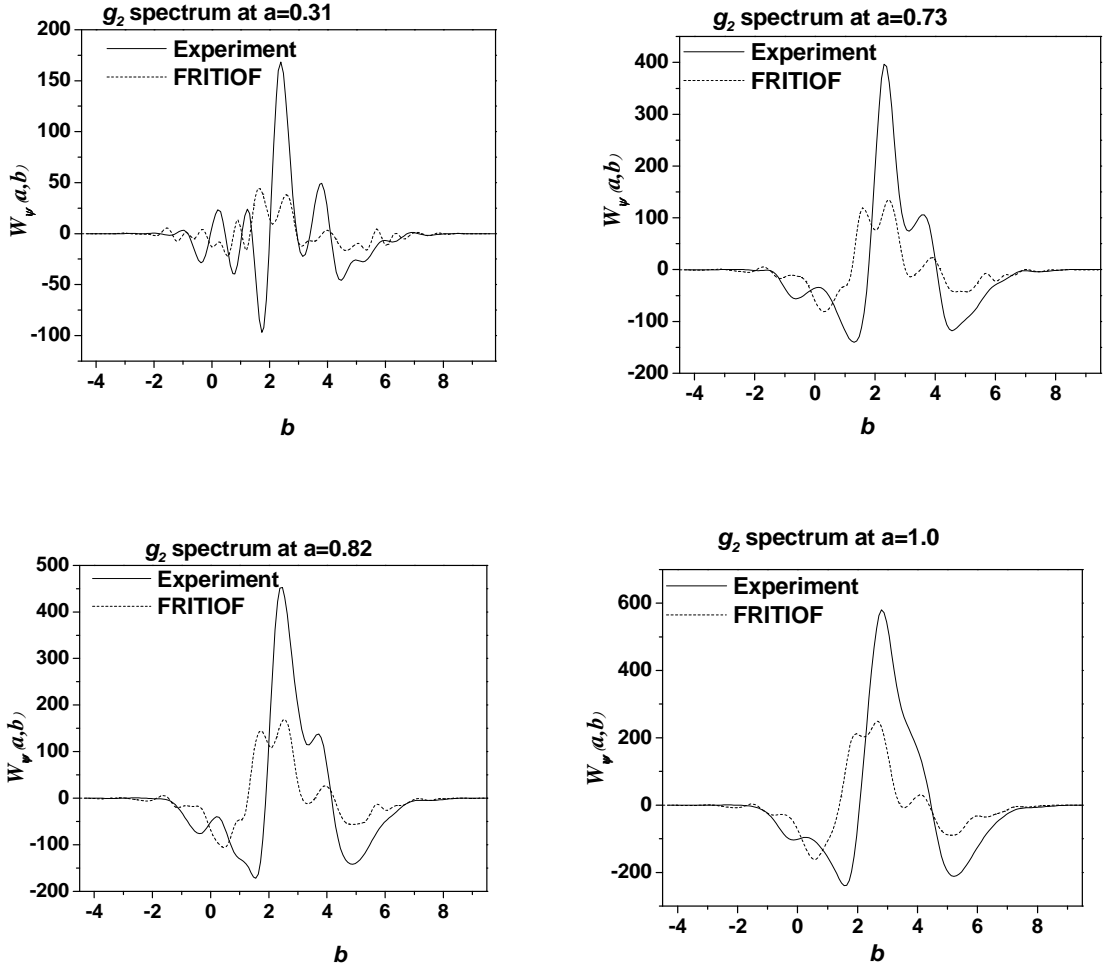

Figure. 2 Wavelet $g_2$ pseudo-rapidity spectra of shower track distribution of $^{16}$O-AgBr interactions at 60 AGeV

Similar plots of $^{32}$S-AgBr interactions at 200 AGeV for entire sample has been shown in figure 3. The broad features of the plots are more or less similar for both interactions. However it is clear that more number of peaks (at least 9) appear in the later one. At small scale $^{32}$S-AgBr interactions are more structured and fluctuating. Due to higher projectile mass and collision energy we observe $^{32}$S-AgBr interactions are more unstable and tumultuous than $^{16}$O-AgBr events which show relatively smooth spectrum. In this case the central particle producing region correspond to $\eta \sim 3.5$, peaks at the left side ($\eta \sim -1.0, 0.1, 1.2$) correspond to target fragmentation region and that at $\eta \sim 4.5, 5.4, 6.2, 7.1, 8.6$ correspond to projectile fragmentation region. Overabundance of produced pions at some preferred pseudo-rapidity may hint towards the presence of ring-like correlation in considered relativistic heavy ion interactions. However for both the interactions we observe that FRITIOF events fail to interpret the experimental results. Fluctuations are much smaller for the simulated events.

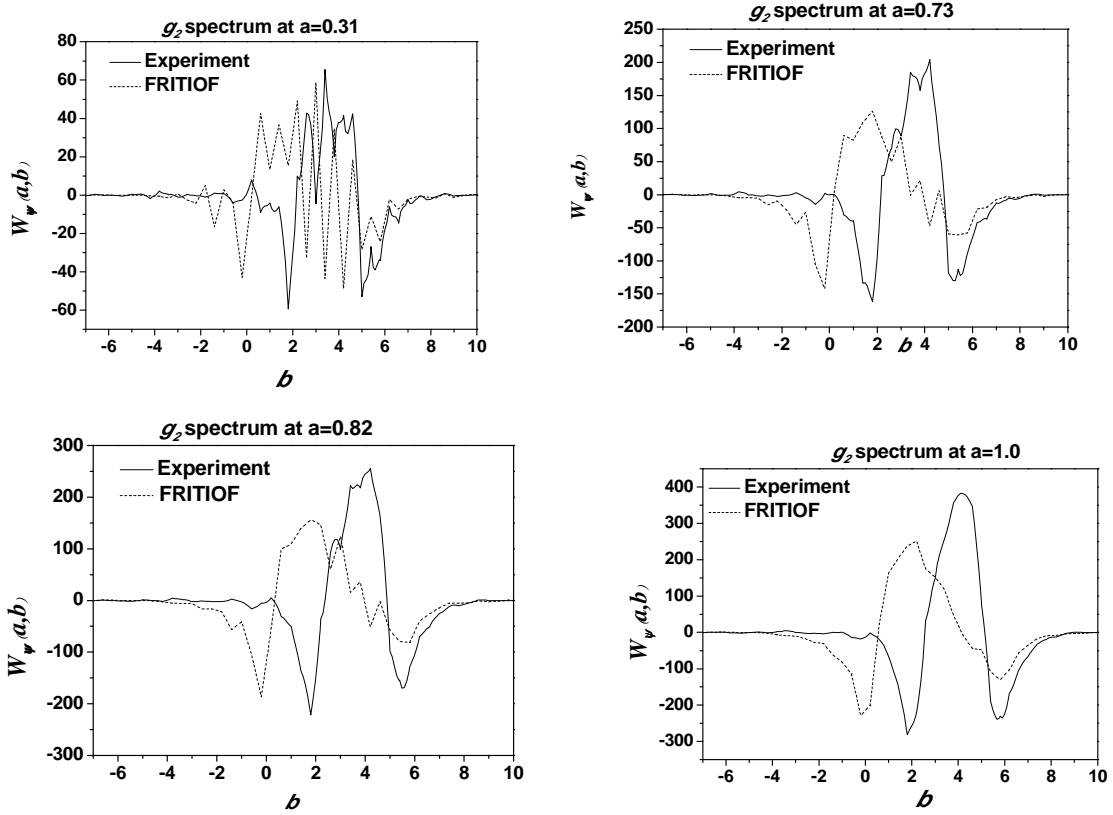

Figure. 3 Wavelet $g_2$ pseudo-rapidity spectra of shower track distribution of $^{32}$S-AgBr interactions at 200 AGeV

It would be interesting to analyse wavelet spectra of individual events of both the interactions using same approach as before to find any characteristic change in particle production of any particular event than in overall process. Two such spectra have been presented in figure 3, for the two highest multiplicity events: one from $^{16}$O-AgBr interactions (multiplicity 114) and other from $^{32}$S-AgBr interactions (multiplicity 229). It has been already pointed out that at very small scales only individual particles are resolved and for larger scales individual particles lose their identity to a bunch. So in these two extreme cases study of clusterization effect is not important. From figure 3 we observe that for a>0.04 several groups of particles are formed.

For the O-event four groups of particles are present at $\eta\sim$ 1.25, 2.15, 3.15, 4 and 5. For the S-event several groups of particles are observed around $\eta\sim$ -1, 0, 2.25, 5, 6, 7 and 8.

From figure 3 we can say that most important scales are from 0.04 to 0.7, as the maximums corresponding to group of particle production are present in this scale region only.

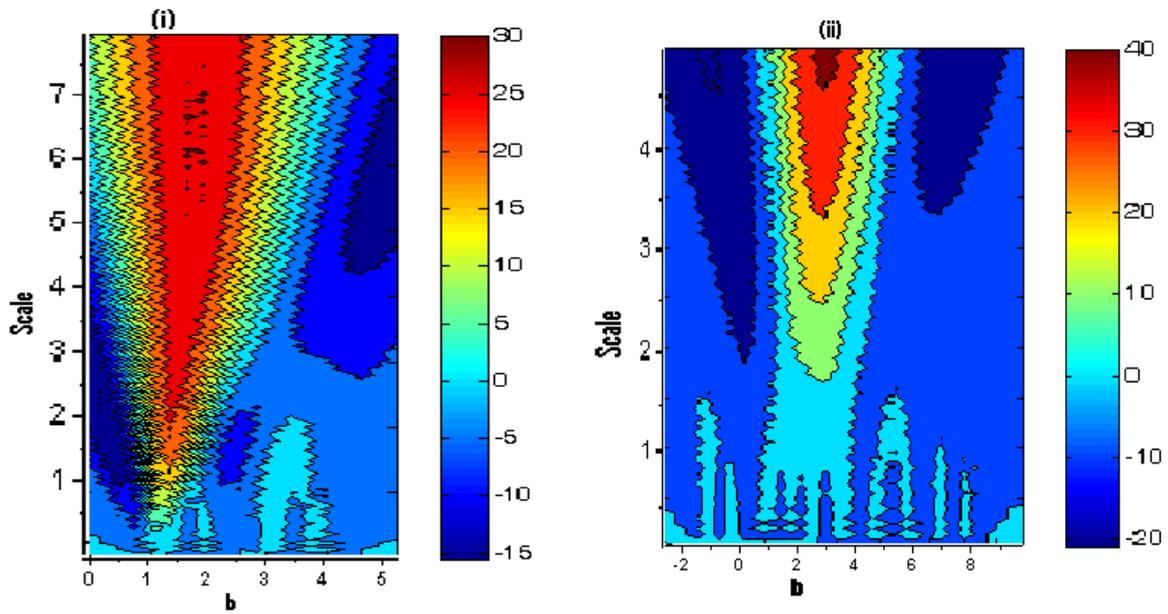

Figure.3 Wavelet pseudo-rapidity spectra for a single event i) for $^{16}$O-AgBr interactions (multiplicity 114), (ii) $^{32}$S-AgBr interactions (multiplicity 229)

For a clear view of most significant scales, it is important to get the scalogram of the corresponding wavelet spectra for the individual events. The scalogram ($E_\omega(a)$), representing the 1-d energy distribution with respect to scale a, is defined as

$$E_\omega(a) = \int \{W_\Psi(a,b)\}^2 \, db \qquad (3)$$

The plots of $E_\omega(a)$ vs. a have been shown in figure 4. Local maximums in these plots represent the dominant scales. Maximums for a<0.05 are insignificant as they results due to statistical reason. From figure 4 (i), the significant scales for O-event are restricted in the range 0.1<a<0.7 and that for S-event is 0.1<a<0.9. In these cases also FRITIOF events are found to be unable to reproduce the experimental findings.

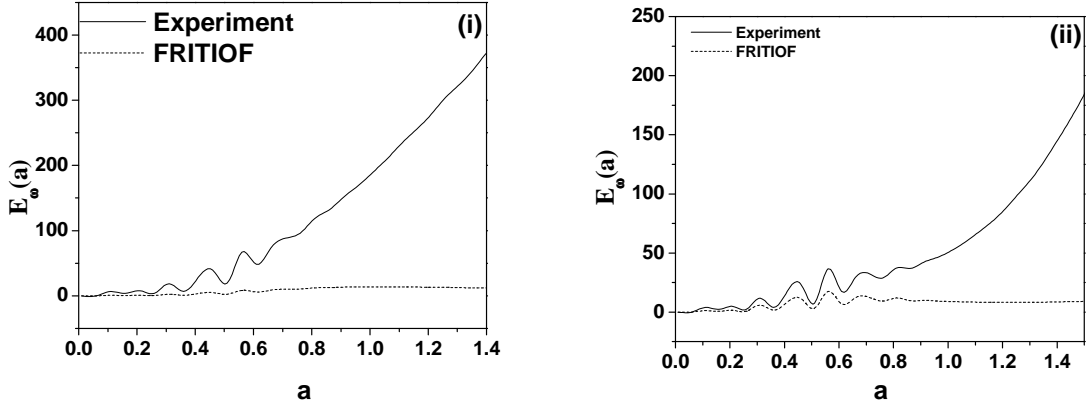

Figure 4. Scalogram of a single event: (i) $^{16}$O-AgBr interactions (multiplicity 114), (ii) $^{32}$S-AgBr interactions (multiplicity 229)

# V. Conclusions

Pseudo-rapidity distributions of pions coming out from $^{16}$O-AgBr & $^{32}$S-AgBr interactions at 60 AGeV & 200 AGeV respectively have been analyzed on the basis of continuous wavelet transformation. The following conclusions may be drawn from the analysis.

- The local maximums, which correspond to the irregularities, are revealed mainly in the scale region a≤0.8.
- The irregularities, found from the multiscale analysis of wavelet pseudorapidity spectra, reveals preferred pseudorapidity values suggests clustering effect in the particle production process.
- Overabundance of particle multiplicity at some distinguished pseudorapidity may be a signature of ring-like correlation.
- Events generated using FRITIOF code fails to reproduce the experimental behaviours exactly, which may be due to the fact that corresponding dynamical input has not been taken into account for developing the code.

## Acknowledgement

The authors are grateful to Prof. P.L. Jain, State University of Buffalo, Buffalo, U.S.A, for providing them with the exposedand developed emulsion plates used for this analysis.

The authors declare that there is no conflict of interest regarding the publication of this paper.